# New high fill-factor triangular micro-lens array fabrication method using UV proximity printing

T.-H. Lin[1], H. Yang[2], C.-K. Chao[1]

[1]Department of Mechanical Engineering, National Taiwan University of Science and Technology, Taipei, Taiwan 106
[2]Institute of Precision Engineering, National Chung Hsing University, Taichung, Taiwan 402

*Abstract*-A simple and effective method to fabricate a high fill-factor triangular microlens array using the proximity printing in lithography process is reported. The technology utilizes the UV proximity printing by controlling a printing gap between the mask and substrate. The designed approximate triangle microlens array pattern can be fabricated the high fill-factor triangular microlens array in photoresist. It is due to the UV light diffraction to deflect away from the aperture edges and produce a certain exposure in photoresist material outside the aperture edges. This method can precisely control the geometric profile of high fill factor triangular microlens array. The experimental results showed that the triangular micro-lens array in photoresist could be formed automatically when the printing gap ranged from 240 μm to 840 μm. The gapless triangular microlens array will be used to increases of luminance for backlight module of liquid crystal displays.

## Ⅰ. INTRODUCTION

The refractive microlens arrays have extensive applications in the optical communication, digital displays, and optical interconnection fields. The integrated microlens array also provides interesting applications for various applications such as enhancing the illumination brightness and simplifying light-guide module construction. In a laptop display, a 25% increase in light output was reported when using this microlens technology (Ezell 2001) [1]. The gapless triangular microlens array can increase 15.1% of luminance in the backlighting module of liquid crystal displays [2]. The microlens shape and the fill-factor are two of many conditions that impact the overall light efficiency. To collect the maximum amount of light, the lens area must be as close to 100% of the total area as possible. The fill-factor is defined as the percentage of lens area to the total area. The fill factor is influenced by the pixel geometry and lens layout [3].

For a round microlens array fabrication, many methods have been reported, including thermal reflow [4], grayscale mask [5], focused-ion-beam (FIB) milling [6], and etching [7]. The fill-factor for a round microlens array was considered in orthogonal and in hexagonally arranged arrays. In an orthogonal array, the maximum fill factor (assuming that no gap between the lenses is necessary) is 78%. In a hexagonally arranged array for a round microlens, the fill-factor is larger (about 90%). Both of these lens configurations have a lower fill-factor than a square, hexagonal and triangular microlens array with a maximum fill factor of nearly 100%. Fig. 1 illustrates a schematic comparison between these high fill factor micro-lens arrays. Each lens has the same dimension of circumscribed circle (encircle) of 78 $\mu$m in radius. It reveals that compared to the other two kinds of microlens arrays, as shown in Fig. 1 (b) and (c), respectively, the layout of triangular microlens array (see Fig. 1(a)) has the largest number of lenses (about 25 lenses) in the same confined area of 500 $\mu$m × 500 $\mu$m. The gapless square and hexagonal micro-lens arrays only show about 18 and 9 lenses, respectively.

Lin *et al* [8] and Yang *et al* [9] presented a special fabrication process that used the incomplete developing technique to produce the hexagonal and square microlens array with a maximum fill factor of nearly 100%. It provides a plastic mold insert for electroforming to convert into a metallic mold for mass production. Furthermore, Lin *et al* proposed a novel high fill factor hexagonal microlens array fabrication method that controlled the printing gap in the UV lithography process [10]. This method can precisely control the geometric profile of the microlens array in the fabrication process without thermal reflow. The cycle time and thermal budget in the fabrication process can therefore be reduced. Pan et al. [2] used a new process to fabricate a high fill factor triangular microlens array. This process mainly includes conventional ultraviolet (UV) lithography, photoresist reflow process, Ni–Co electroplating. In this study, the UV proximity printing followed by a photoresist is proposed here to provide a simple method of fabricating high fill factor triangular microlens arrays. The proposed method can precisely control the geometric profile of triangular microlens array. This work also offers the novel and simple fabrication method for high fill-factor microlens array. The application of this microlens array can increase the luminance in the backlight module of liquid crystal displays.

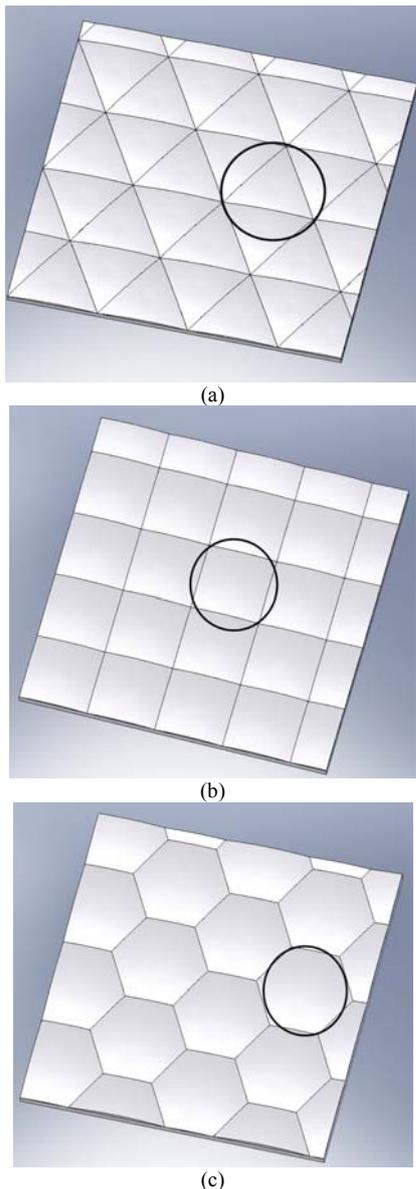

Fig. 1. Schematic illustration of lens number in the same area of 500μm×500μm: (a) gapless triangular micro-lens arrays with 25 lenses; (b) gapless square micro-lens arrays with 18 lenses; (c) gapless hexagonal microlens arrays with 9 lenses.

## II. EXPOSURE CHARACTERISTICS

The lithographic exposure has three printing modes including contact, proximity and projection. The exposure operations using the proximity mode occur in the near field or Fresnel diffraction regime. The pattern resulting from the light passing through the mask directly impacts the photoresist surface because there is no lens between the photoresist and mask. The created aerial image therefore depends on the near field diffraction pattern. Because of the diffraction effects, the light bends away from the aperture edges and produces partial exposure outside the aperture edges. Although the contact mode can minimize these effects by reducing the gap to zero, the gap is not strictly zero in practice because the top surface of the photoresist is not perfectly flat.

Fig. 2 shows the intensity distribution in the mask and photoresist as the light passes through the apertures. The middle figure shows a regular light intensity profile in mask. The smooth and convex surface profile is formed in photoresist as shown in the bottom figure. The convex profile can be fabricated through proper operational parameters using a negative photoresist. The desired microstructures are formed after the development process. As the gap spacing increases between the mask and photoresist, the aerial image quality produced on the photoresist surface will strongly degrade due to the diffraction effects [11]. That is, the intensity distribution range becomes wider and the relative light intensity becomes weaker at some point for one aperture exposure. The aperture exposure arranged using the array produces a complicated three-dimensional light intensity distribution in the photoresist. The final concave micro-structure geometry can be determined using the resulting intensity distribution after exposure and development.

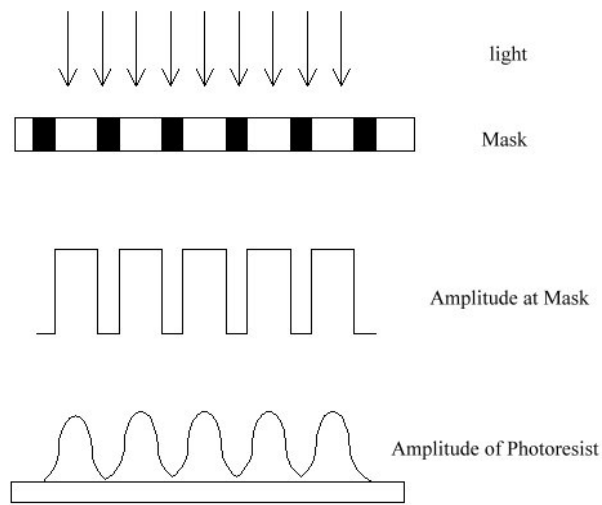

Fig. 2. Light intensity distribution in photoresist exposed using two adjacent apertures

## III. EXPERIMENT

Desired patterns are transferred from the designed mask in lithographic process. In this experiment, a plastic mask was fabricated using laser writing onto a PET (Polyethylene terephthalate) used for PCBs (print circuit boards). The layout pattern on the plastic mask is illustrated in Fig. 3. The distance between of the triangle vertices and circumcenter was L=80μm in the each aperture and the pitch distance for two adjacent middle point side of apertures was p=0μm. The upper and lower rows were arranged in equidistance. Polycarbonate (PC) sheets 2 mm thick were used as the substrate. To increase the adhesive force between the photoresist and substrate, the PC substrate was coated first with a thin layer of HMDS. The PC substrate was then spun with a layer of negative photoresist (JSR THB-120N) 35 lm thick. The spin condition was 1100 rpm for 20 sec. Prebaking in a convection oven at 90°C for 3 min is a required procedure. This removes the excess solvent from the photoresist and produces a slightly hardened photoresist surface. The mask was not stuck onto the substrate. The sample was exposed through the plastic mask using a UV mask aligner (EVG620). This aligner had soft, hard contact, or proximity exposure modes with near ultra-violet wavelength 350–450 nm and lamp power range from 200 to 500 W. A slice of glass was inserted between the photoresist and mask to create a gap. Exposure was then conducted for about 16 sec. The exposure process is shown in Fig. 4. After exposure, dipping into the developer for 3 min and cleaning with deionized water, the 3-D array was completed. The gap was adjusted to 120μm, 240μm, 360μm, 480μm, 600μm, 720, 840, 960μm. The triangular microlens array variations in the samples were observed and the optimal gap range determined. The scanning electron microscopy (SEM) and 3D surface profile were used to measure the characteristics of the resulting microlens array structures.

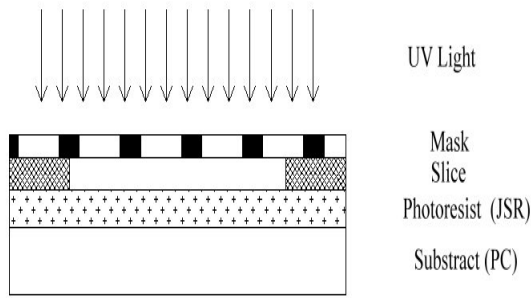

Fig. 3. Schematic of the high fill-factor triangular microlens array fabrication using UV proximity printing

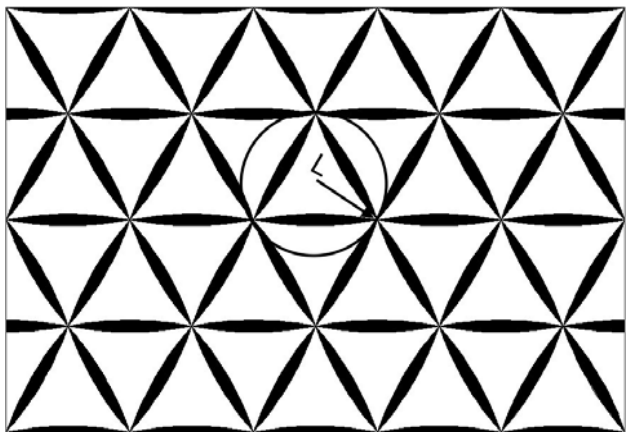

Fig. 4. Illustration of the pattern design on the mask for high fill factor triangular microlens array fabrication, the upper and lower rows of equidistance were designed. (Unit: μm)

## IV. RESULTS

According to the experimental results, three high fill-factor triangular microlens arrays were classified after exposure by using different gap sizes. The printing gap sizes were ranged from 120 μm to 960 μm. Firstly, a flat down microlens array mold was formed using the printing gap 120 μm. Fig. 5 shows the flat top structural array by using SEM observation. Secondly, adjusting the printing gap between 240μm to 840μm, the high fill-factor triangular microlens array was formed as shown in Fig. 6 and Fig.7. Thirdly, using the printing gap is larger than 840 μm, the top view is blurred. The printing gap between the photoresist and mask gradually increased, the height and sternness in the high fill-factor triangular microlens array at the lateral sideline became thinner and flatter. A small printing gap (less than 240 μm) is not suitable for high fill-factor triangular microlens array fabrication because the light intensity distribution in the photoresist is not have enough diffraction to produce convex spherical structures. The cross-sectional and 3-D surface profiler high fill-factor triangular microlens array was measured using a 3-D surface profiler, as shown in Fig. 8. Its actual profile compared to the theoretical spherical curve (dashed line) of a spherical lens is depicted in Fig. 9. It shows that the deviation on the lens sag from a perfect sphere is quite small. The high fill factor triangular microlens array surface roughness was determined using an atomic force microscope (AFM). The measured size was $5 \times 5 \mu m^2$ on the triangular microlens top. It shows that the microlens surface roughness (Ra) was 5.77nm in Fig. 10.

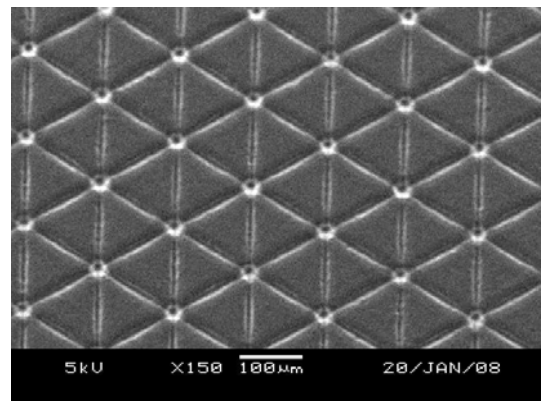

Fig. 5. SEM micrograph of the high fill-factor triangular microlens array fabricated using the printing gap sizes less than 240.

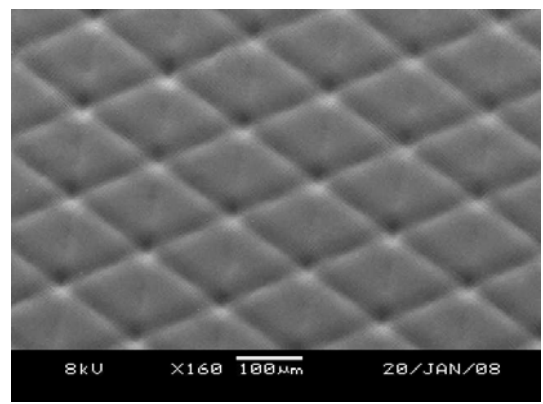

Fig. 6. SEM micrograph of the high fill-factor triangular microlens array fabricated using a 240 μm printing gap.

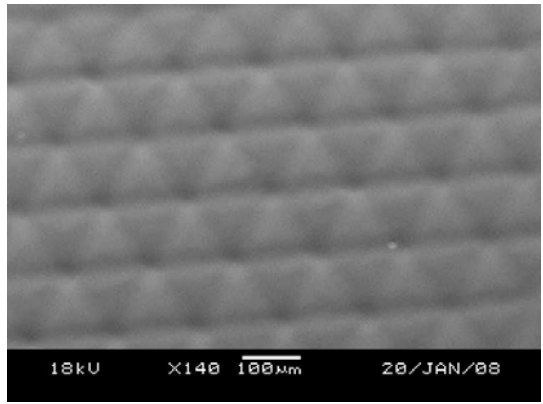

Fig. 7. SEM micrograph of the high fill-factor triangular microlens array fabricated using a 840 μm printing gap.

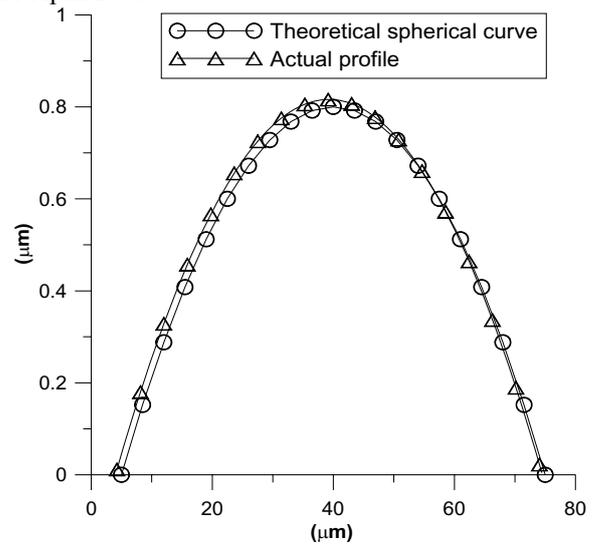

Fig. 9. Cross-sectional profile of a triangular micolens as measured with the stylus profilmeter.

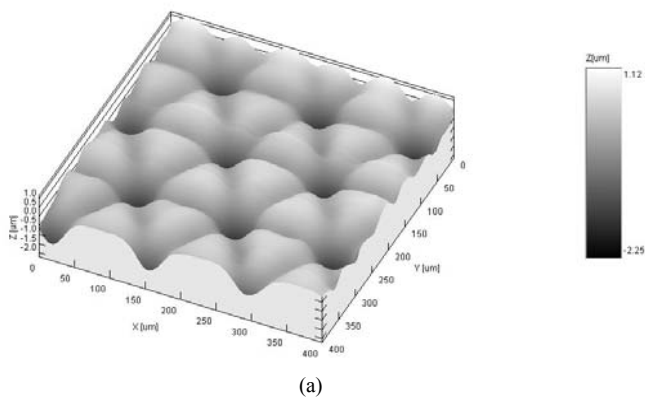

(a)

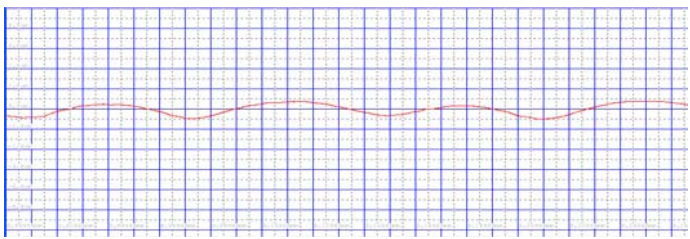

(b)

Fig. 8. Experimental results of the high fill-factor triangular microlens array; (a) three-dimensional profile measurement, and (b) cross-sectional profile of the high fill-factor triangular microlens array.

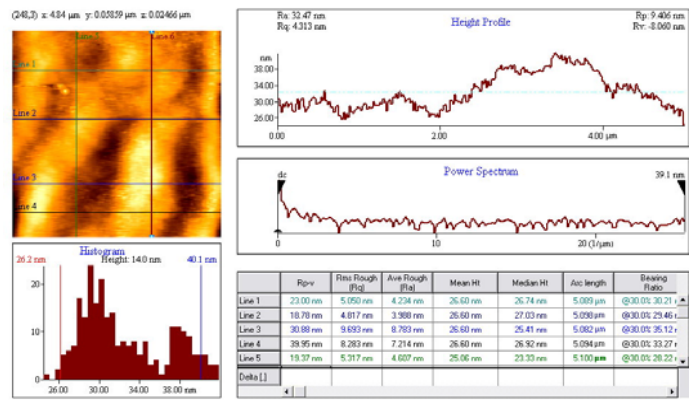

Fig. 10. The high fill factor triangular microlens surface roughness using the AFM measurement result.

## V. CONCLUSION

The fabrication method of the high fill-factor triangular microlens array is presented. The experimental results showed that the high fill-factor triangular micro-lens array in photoresist can be formed automatically when the printing gap ranged from 240 μm to 840 μm. The proposed method can precisely control the geometric profile of triangular microlens array. This work also offers the novel and simple fabrication method for high fill-factor microlens array. The application of this microlens array can increase the luminance in the backlight module of liquid crystal displays.

ACKNOWLEDGMENT

This research is supported by the National Science Council of Taiwan under the grand number NSC96-2221-E-005-069-MY3.